\documentclass[%
 aip,
amsmath,amssymb,
preprint,%
]{revtex4-2}

\usepackage{dcolumn}
\usepackage[mathlines]{lineno}

\usepackage{afterpage}
\usepackage{longtable}
\usepackage{hyperref}
\usepackage[english]{babel}
\usepackage{dcolumn}
\usepackage{bm}
\usepackage{pifont}
\usepackage{amsfonts}
\usepackage{amsmath}
\usepackage{graphicx}
\usepackage{natbib}
\usepackage{psfrag}
\usepackage{color}
\usepackage{multirow}
\usepackage{cases}
\usepackage{stackengine}
\usepackage{float}
\usepackage{blkarray}
\usepackage{cancel}
\usepackage{mathrsfs}
\usepackage{empheq}
\usepackage[utf8]{inputenc}
\usepackage[T1]{fontenc}
\usepackage{mathptmx}

\newcommand*\patchAmsMathEnvironmentForLineno[1]{%
  \expandafter\let\csname old#1\expandafter\endcsname\csname #1\endcsname
  \expandafter\let\csname oldend#1\expandafter\endcsname\csname end#1\endcsname
  \renewenvironment{#1}%
     {\linenomath\csname old#1\endcsname}%
     {\csname oldend#1\endcsname\endlinenomath}}%
\newcommand*\patchBothAmsMathEnvironmentsForLineno[1]{%
  \patchAmsMathEnvironmentForLineno{#1}%
  \patchAmsMathEnvironmentForLineno{#1*}}%
\AtBeginDocument{%
\patchBothAmsMathEnvironmentsForLineno{equation}%
\patchBothAmsMathEnvironmentsForLineno{align}%
\patchBothAmsMathEnvironmentsForLineno{flalign}%
\patchBothAmsMathEnvironmentsForLineno{alignat}%
\patchBothAmsMathEnvironmentsForLineno{gather}%
\patchBothAmsMathEnvironmentsForLineno{multline}%
}

\providecommand\bnabla{\boldsymbol{\nabla}}
\providecommand\bcdot{\boldsymbol{\cdot}}
\providecommand\upi{\pi}%

\newcommand{\V}{\mathcal{V}}
\newcommand{\Surf}{\partial \V}

\newcommand{\Ca}{\mbox{\textit{Ca}}} 	        

\DeclareMathAlphabet{\mathsfbi}{\encodingdefault}{\sfdefault}{bx}{sl}

\newcommand{\hmin}{h_\textrm{min}}

\newcommand{\tr}{t_\textrm{R}}

\newcommand{\com}[1]{}

\begin{document}

\preprint{Preprint accepted in Phys. Fluids}

\title{The influence of an outer bath on the dewetting of an ultrathin liquid film}

\author{A. Mart\'inez-Calvo}
\affiliation{Princeton Center for Theoretical Science, Princeton University, Princeton, NJ 08544, USA.}
\affiliation{Department of Chemical and Biological Engineering, Princeton University, Princeton, NJ 08544, USA.}

\author{D. Moreno-Boza}
\email{damoreno@pa.uc3m.es}
\affiliation{Departamento de Ingenier\'ia T\'ermica y de Fluidos, Universidad Carlos III de Madrid. Avda. de la Universidad 30, 28911, Legan\'es, Madrid, Spain.}

\author{J. F. Guil-Pedrosa}
\affiliation{Departamento de Ingenier\'ia T\'ermica y de Fluidos, Universidad Carlos III de Madrid. Avda. de la Universidad 30, 28911, Legan\'es, Madrid, Spain.}

\author{A. Sevilla}
\affiliation{Departamento de Ingenier\'ia T\'ermica y de Fluidos, Universidad Carlos III de Madrid. Avda. de la Universidad 30, 28911, Legan\'es, Madrid, Spain.}

\begin{abstract}
We report a theoretical and numerical investigation of the linear and nonlinear dynamics of a thin liquid film of viscosity $\mu$ sandwiched between a solid substrate and an unbounded liquid bath of viscosity $\lambda \mu$. In the limit of negligible inertia, the flow depends on two non-dimensional parameters, namely $\lambda$ and a dimensionless measure of the relative strengths of the stabilizing surface tension force and the destabilizing van der Waals force between the substrate and the film. We first analyze the linear stability of the film, providing an analytical dispersion relation. When the viscosity of the outer bath is much larger than that of the film, $\lambda \gg 1$, the most amplified wavenumber decreases as $k_{\textrm{m}} \sim \lambda^{-1/3}$, indicating that very slender dewetting structures are expected when $\lambda$ becomes large. We then perform fully nonlinear simulations of the complete Stokes equations to investigate the spatial structure of the flow close to rupture revealing that the flow becomes self-similar with the minimum film thickness scaling as $\hmin = K(\lambda) \tau^{1/3}$ when $\tau \to 0$, where $\tau$ is the time remaining before the singularity. It is demonstrated that the presence of an outer liquid bath affects the self-similar structure obtained by~\cite{morenobozaetal2020} through the prefactor of the film thinning law, $K(\lambda)$, and the opening angle of the self-similar film shape, which is shown to decrease with $\lambda$. 
\end{abstract}

\date{\today}

\maketitle

\section{Introduction\label{sec:intro}}

Liquid films are encountered across scales and material properties, be it in nature or in a large number of technological applications, ranging from rheologically complex geological flows, tear films, or wine legs, to nanometric metal coatings in plasmonic devices. For a panoramic view of the field, the reader is referred to the reviews~\cite{Oron1997},~\cite{craster2009dynamics}, and~\cite{kondic2019liquid}. A configuration that is routinely used for the fabrication of such nanostructures, is that of a viscous ultra-thin liquid film resting on a substrate which is embedded in a viscous outer bath. The control of these films is crucial to obtain the final product with a homogeneous coating, which requires a fundamental understanding of the forces involved in their stability and nonlinear dynamics. 

This paper is devoted to study the linear and nonlinear regimes of this particular configuration when the film is unstable to van der Waals (vdW) forces between the solid substrate and the film, a destabilization process that typically occurs when the thickness of the film drops below 100 nm approximately. Below this thickness, the vdW force overcomes the stabilizing surface tension force, which triggers a self-accelerated process leading to the spinodal dewetting process. As the thickness of the film vanishes, a finite-time singularity arises in the equations of motion, where the pressure, stresses and velocity diverge as rupture is approached. These nonlinear near-rupture dynamics have been extensively investigated when the film is embedded in a passive ambient, by means of lubrication theory~\citep{ZhangLister1999} and also considering the full conservation equations~\citep{morenobozaetal2020,moreno2020inertial}. Close to the singularity, the structure of the flow is self-similar and universal, in that it does not depend on the initial or boundary conditions and lacks from a characteristic length scale. 

In our previous work~\citep{morenobozaetal2020}, we showed that the self-similar thinning of the liquid film predicted by lubrication theory is only observed during a transient that depends on the initial height of the film, but it is not established as rupture is approached and the slenderness of the flow breaks down. The complete Stokes equations predict a non-slender universal structure of the flow associated with a film shape that exhibits an opening angle of 37$^{\textrm{o}}$ with respect to the substrate. In this work we extend the results of~\cite{morenobozaetal2020} and study the dynamics of an ultra-thin liquid film embedded in a viscous outer bath.

Flow configurations related with the one reported herein have been studied in the context of the dewetting of two-layer thin films~\citep{pototsky2004alternative,fisher2005nonlinear,merkt2005long,nepomnyashchy2009instabilities}, although the finite thickness of the outer layer has an essential effect on the dynamics. A similar flow also takes place during the late stages prior to drop coalescence, which are locally driven by vdW interactions close to contact~\citep{beaty2022nonuniversal,beaty2023inertial}. \com{Also, the linear dynamics of a thin liquid layer surrounded by an outer bath have also been looked at recently in~\cite{mirjalili2021linear}}. However, to the best of our knowledge, the dynamics of a thin film sandwiched between a solid substrate and an unbounded liquid bath has not been reported before, and therefore constitutes the main objective of the present contribution.

The paper is structured as follows. In section~\S\ref{sec:eqs} we detail the governing equations of motion, which are then used in section~\S\ref{sec:lsa} to study the linear stability of the film, and in section~\S\ref{sec:rupture} to perform direct numerical simulations. In particular, in the latter section we investigate the structure of the flow near rupture and the existence of self-similarity. Conclusions are finally drawn in~\S\ref{sec:conclusions}.

\begin{figure}
    \centering
    \includegraphics[width=450pt]{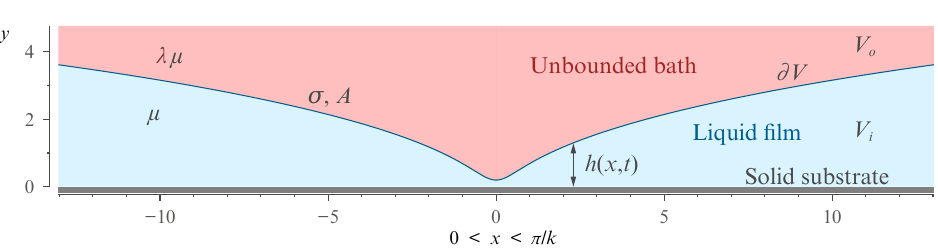}
    \caption{Sketch of the flow configuration and controlling parameters.}
    \label{fig:fig1}
\end{figure}

\section{Equations of motion}\label{sec:eqs}
We consider the Newtonian flow of a thin liquid film of viscosity $\mu$ and initial height $h^{*}_{\textrm{o}}$, resting on a solid substrate and immersed in a bath of viscosity $\lambda \mu$ (see sketch in figure~\ref{fig:fig1}), with $\sigma$ denoting the surface tension coefficient. To describe the flow we make use of the incompressible Stokes equations, which are non-dimensionalized with the following characteristic length, velocity, time, pressure, and intermolecular potential scales, respectively,
\begin{equation}\label{eq:scales}
\ell_{\textrm{c}} = a = \sqrt{\frac{A}{6\upi \sigma}}, \quad v_{\textrm{c}} = \frac{\sigma}{\mu}, \quad t_{\textrm{c}} = \frac{\mu a}{\sigma}, \quad p_{\textrm{c}} = \phi_{\textrm{c}} = \frac{A}{6 \upi a^3},
\end{equation}
where $a$ is the molecular length scale~\citep{de1985wetting,ZhangLister1999,morenobozaetal2020} and $A$ the Hamaker constant~\citep{hamaker1937london}, to yield
\begin{equation}\label{eq:ns}
\bnabla \bcdot \bm{v}_i = 0, \quad \text{and} \quad \bm{0} =  \bnabla \bcdot \bm{T}_i \quad \bm{x} \in \V_i,
\end{equation}
where $i = \{\textrm{in},\textrm{out}\}$ denotes inner and outer bath fluid, respectively, $\bm{T}_{\textrm{in}} = -p_{\textrm{in}} \bm{I} + \bnabla \bm{v}_{\textrm{in}} + \bnabla \bm{v}_{\textrm{in}}^{\textrm{T}}$ and $\bm{T}_{\textrm{out}} = -p_{\textrm{out}} \bm{I} + \lambda \left(\bnabla \bm{v}_{\textrm{out}} + \bnabla \bm{v}_{\textrm{out}}^{\textrm{T}}\right)$ are the stress tensors, $\bm{I}$ is the identity tensor, and $p$ is the pressure field. Here $\V_i$ denotes the liquid domains, $\bm{x} = (x,y)$ is the positional vector field, and $\bm{v} = (u,v)$ is the velocity field, assumed to be two-dimensional. At the free surface $\Surf$ we impose the continuity of velocities, the kinematic and stress balance boundary conditions, which read
\begin{subequations}\label{eq:interface}
\begin{alignat}{2}
& \bm{v}_{\textrm{in}} = \bm{v}_{\textrm{out}}, \quad \bm{x} \in \Surf \label{eq:continuity_vel}\\
&\bm{n}\bcdot\left( \partial_t \bm{x}_{\textrm{s}} - \bm{v} \right) = 0, \quad \bm{x} \in \Surf \label{eq:interface_kinematic} \\
&\left(\bm{T}_{\textrm{in}} - \bm{T}_{\textrm{out}} + h^{-3} \bm{I}\right) \bcdot \bm{n} = - \bm{n} (\bnabla \bcdot \bm{n}),  \quad \bm{x} \in \Surf \label{eq:interface_stress}
\end{alignat}
\end{subequations} 
respectively, where $\bm{x}_{\textrm{s}}$ is the parametrization of the interface, located at $y = h(x,t)$, and $\bm{n}$ is the unit normal vector to the interface. At the solid substrate, $y = 0$, the no-slip boundary condition is enforced, $\bm{v}_{\textrm{in}} = \bm{0}$, and at $y \to \infty$ we impose $\bm{T}_{\textrm{out}} \bcdot \bm{e}_y = \bm{0}$~\com{, where $\bm{e}_y$ is the unit vector along the vertical direction}. As for the initial conditions, in the numerical simulations we consider half a wavelength of a spatially periodic liquid film, and thus we impose the symmetry condition $u_\mathrm{in} = 0$ at $x = 0$ and $x = \upi/k$, where $k < k_{\textrm{c}} = \sqrt{3}/h_o$ is the dimensionless wavenumber of the initially perturbed interface $\bm{x}_{\textrm{s}} = \left[x, h_{\textrm{o}} \, (1-\epsilon \cos{kx})\right]$, imposed at $t = 0$. Here, $k_{\textrm{c}}$ is the dimensionless cut-off wavenumber predicted by linear instability theory~\citep{Vrij1966}, $h_{\textrm{o}} = h^{*}_{\textrm{o}}/a$ is the initial film thickness normalised with the molecular length scale, and $\epsilon$ is a small constant that triggers the instability and induces the rupture of the liquid film at $x=0$ and $t=\tr$.

Equations~\eqref{eq:ns}--\eqref{eq:interface} can be rewritten in terms of the alternative scalings 
\begin{equation}\label{eq:scales_ho}
\ell_{\textrm{c}} = h^{*}_{\textrm{o}}, \quad v_{\textrm{c}} = \frac{A}{6 \pi \mu {h^{*}_{\textrm{o}}}^2}, \quad t_{\textrm{c}} = \frac{6 \pi \mu  {h^{*}_{\textrm{o}}}^3}{A}, \quad p_{\textrm{c}} = \phi_{\textrm{c}} = \frac{A}{6 \upi {h^{*}_{\textrm{o}}}^3},
\end{equation} 
based on the initial film thickness $h_{\textrm{o}}^*$, which are suitable for the analysis of the linear stage of the film evolution. The capillary number $\Ca = A/(6 \pi \sigma {h_{\textrm{o}}^*}^2) = h_{\textrm{o}}^{-2}$ comparing vdW-effects to surface tension is seen as the controlling parameter using the scales given in~\eqref{eq:scales_ho}.

\section{Linear stability analysis}\label{sec:lsa}
We first investigate the linear stability of the film by deducing the growth rate of disturbances from the complete Stokes equations~\eqref{eq:ns}--\eqref{eq:interface_stress}. To this end, by making use of the scales given in~\eqref{eq:scales_ho}, we expand all the flow variables~\com{into} normal modes of the form 
\com{
\begin{equation}
    \label{eq:normal_modes}
    (\bm{v}_\mathrm{in,out},p_\mathrm{in,out},h) = (0,p_o,1) + (\tilde{\bm{v}}_\mathrm{in,out},\tilde{p}_\mathrm{in,out},\tilde{h})\exp(ikx + \omega t)
\end{equation}}where~\com{ $p_o$ is the initial uniform pressure,} $\omega$ is the growth rate of perturbations,~\com{ indicated with tildes}, and $k$ the associated axial wavenumber, and solve the corresponding eigenvalue problem which yields the following dispersion relation
\begin{equation}\label{eq:complete_dr}
\omega = \frac{(3 -\Ca^{-1}k^2) [\lambda (1-\cosh(2k))-\sinh(2k) +2k(1+k\lambda)]}{2k[(1+2k^2)(\lambda^2-1) -(1+\lambda^2)\cosh(2k) -2 \lambda \sinh(2k)]}.
\end{equation}
In the limit $\lambda \to 0$ we recover the well-known result of a thin film surrounded by a solid substrate and a passive gaseous ambient~\citep{jain1976stability,morenobozaetal2020}.
\begin{figure}
    \centering
    \includegraphics[width=0.9\textwidth]{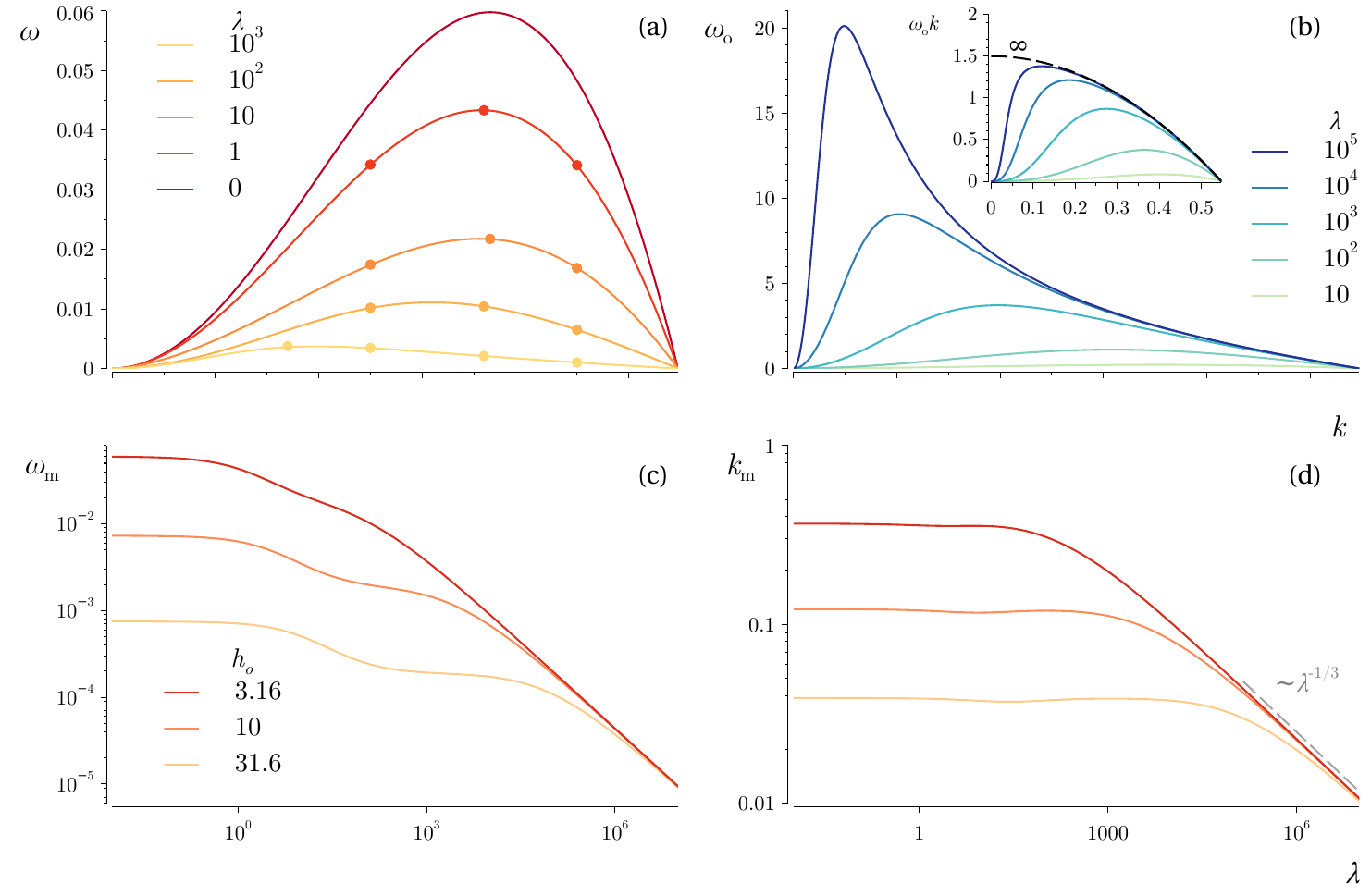}
    \caption{Amplification curves ($a$) $\omega(k)$~\com{ (where dots indicate growth rates obtained from numerical simulations)}, and ($b$) $\omega_{\textrm{o}}(k)=\lambda\omega(k)$ for $h_{\textrm{o}} = 3.16$ ($Ca = 0.1$) and different values of the viscosity ratio $\lambda$ indicated in the legend. The inset in ($b$) shows the function $k\omega_{\textrm{o}}(k)$. ($c$) Maximum growth rate $\omega_{\textrm{m}}$ and ($d$) associated wavenumber $k_{\textrm{m}}$, as functions of $\lambda$ for different values of $h_{\textrm{o}}$ indicated in the legend.}
    \label{fig:fig2}
\end{figure}
To investigate the limit of a passive ultra-thin gaseous film confined by a solid substrate below and by a viscous bath above, corresponding to the limit $\lambda \gg 1$, it is convenient to first re-scale the growth rate using the viscosity of the outer bath, $\omega_{\textrm{o}} = \omega \lambda$. Taking the limit $\lambda \to \infty$ in the dispersion relation~\eqref{eq:complete_dr}, we find
\begin{equation}\label{eq:lambdainf_dr}
\omega_{\textrm{o}} = \frac{3-\Ca^{-1}k^2}{2k}
\end{equation}
for $k\ll 1$, displaying a singular growth rate for $k\to 0$. A common feature of inertialess unstable interfacial flows that lack from shear associated with a solid boundary is that the most unstable wavenumber is $k = 0$, as happens, for instance, in cylindrical liquid threads~\citep{Rayleigh4} and free liquid films~\citep{erneux1993nonlinear,sharma1995nonlinear}. Here, in addition, the absence of a transverse length scale within the bath makes the growth rate diverge as $k\to 0$. 

Figure~\ref{fig:fig2} summarizes the main results obtained from the linear stability analysis. The temporal growth rate scaled with the film viscosity, $\omega(k)$, is seen to decrease monotonically as $\lambda$ increases, as does the most dangerous wavenumber. In particular, as deduced from Fig.~\ref{fig:fig2}(d), the wavenumber of maximum growth rate $k_{\textrm{m}}\sim \lambda^{-1/3}$ for $\lambda\to \infty$, irrespective of the value of $h_{\textrm{o}}$. In view of the results of the linear stability analysis it is expected that the longitudinal length scale of the dewetting structures increases with $\lambda$, an aspect that will be examined next by means of numerical simulations and similarity theory.

\section{Nonlinear thinning}\label{sec:rupture}
In this section we study the nonlinear and near-rupture dynamics of the thin liquid film. To that end, it proves convenient to take back the original scales~\eqref{eq:scales}, that will be assumed hereafter. First we perform numerical simulations of the full conservation equations~\eqref{eq:ns}--\eqref{eq:interface} in~\S\ref{subsec:numerics}, and then we investigate the self-similar structure of the flow close to the singularity in~\S\ref{subsec:selfsimilar}.

\subsection{Description of the flow evolution}\label{subsec:numerics}

\begin{figure}
    \centering
    \includegraphics[width=\textwidth]{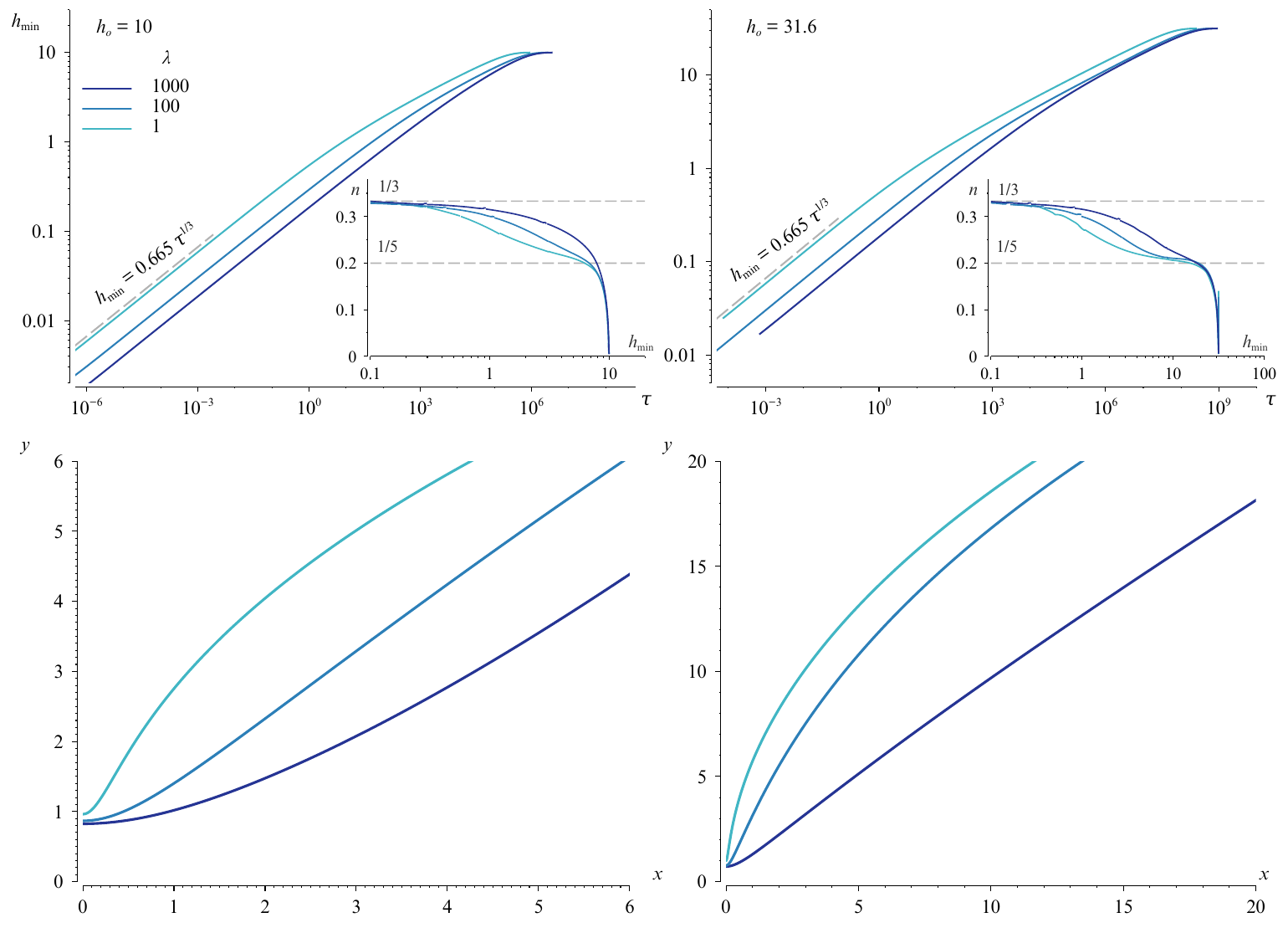}
    \caption{Top panels: minimum film thickness, $\hmin$, as a function of the time to rupture, $\tau$, for $h_{\textrm{o}} = 10$ (left) and $h_{\textrm{o}} = 31.6$ (right), and different values of the viscosity ratio, $\lambda$, indicated in the legend. The insets show the instantaneous power-law exponents, $n={\rm d}\log(\hmin)/{\rm d}\log(\tau)$, as functions of $\hmin$. Bottom panels: corresponding free-surface shapes for $\hmin\approx 1$.}
    \label{fig:fig3}
\end{figure}

The temporal evolution of the unstable film was obtained by integrating the set of equations~\eqref{eq:ns}-\eqref{eq:interface_stress} using the same finite-element technique explained in detail by~\citet{morenobozaetal2020}. To that end, a small-amplitude harmonic disturbance with optimal wavenumber is imposed as initial condition, and evolved in time up to times close to the instant of the rupture singularity, $\tr$. Figure~\ref{fig:fig3} shows the results obtained from simulations performed for two different values of the initial film height, $h_{\textrm{o}}=10$ (left) and $h_{\textrm{o}}=31.6$ (right), and $\lambda=(1,100,1000)$ as indicated in the legend. The upper panels show the minimum film thickness, $\hmin$, as a function of the time to rupture, $\tau=\tr-t$. The simulations show that $\hmin\propto \tau^{1/3}$ as $\tau\to 0$ in all cases, revealing that the same power-law exponent found by~\citet{morenobozaetal2020} for the case without outer bath, $\lambda\to 0$, is actually valid for any value of $\lambda$. The viscosity ratio $\lambda$ is seen to only affect the prefactor of the thinning law, such that $\hmin=K(\lambda)\tau^{1/3}$. The function $K(\lambda)$, represented in Fig.~\ref{fig:fig4}, has the limit $K\to 0.66$ as $\lambda\to 0$, corresponding to a negligible influence of the outer bath~\cite{morenobozaetal2020}, and decreases monotonically with $\lambda$, reaching a power-law behavior $K\propto \lambda^{-2/9}$ for $\lambda\gg 1$. The instantaneous power-law exponents $n={\rm d}\log(\hmin)/{\rm d}\log(\tau)$, represented as function of $\hmin$ in the insets, are seen to monotonically increase towards the $1/3$ power law, which is fully established for values of $\hmin \approx 1$, which correspond to scales close to the molecular one. In the case with larger initial thickness, $h_{\textrm{o}}=31.6$, the $1/5$ power law obtained by~\citet{ZhangLister1999} using lubrication theory for the case $\lambda\to 0$, is established during a brief stage in the cases with $\lambda=1$ and $\lambda=100$. 

Besides its influence on the temporal scaling law for the thinning process, the viscosity ratio $\lambda$ is also seen to profoundly affect the geometry of the interface near rupture. Indeed, the lower panels of Fig.~\ref{fig:fig3} display the free-surface shapes corresponding to the same cases of the upper panels, evaluated at $\hmin\approx 1$. The typical slope of the interface is seen to decrease with $\lambda$, revealing that the flow becomes more slender as the viscosity ratio is increased as anticipated by the linear stability theory of the previous section. In the next section, the actual flow structure close to the rupture singularity will be deduced by means of a similarity analysis.


\subsection{The self-similar flow close to rupture}\label{subsec:selfsimilar}

Following our previous work~\citep{morenobozaetal2020}, dimensional analysis suggests the similarity Ansatz for the near-rupture flow:
\begin{subequations}
\begin{gather}
x = \tau^{1/3} \xi, \quad y = \tau^{1/3}\eta, \quad h = \tau^{1/3}f(\xi), \quad u_{i} = \tau^{-2/3}U^{i}(\xi,\eta), \label{eq:ss_scales1} \\
v_{i} = \tau^{-2/3}V^{i}(\xi,\eta), \quad p_{i} = \tau^{-1} P^{i}(\xi,\eta), \label{eq:ss_scales2}
\end{gather}
\end{subequations}
which when introduced into Eqs.~\eqref{eq:ns} yields, at leading order in $\tau$, the set of equations
\begin{subequations}
\begin{gather}
\begin{rcases} 
      U^{\textrm{in}}_{\xi} + V^{\textrm{in}}_{\eta} = 0, \\
      U^{\textrm{in}}_{\xi \xi} + U^{\textrm{in}}_{\eta \eta} = P^{\textrm{in}}_{\xi} - 3 f^{-4} f_{\xi},\\
      V^{\textrm{in}}_{\xi \xi} + V^{\textrm{in}}_{\eta \eta} = P^{\textrm{in}}_{\eta},
\end{rcases} \quad \text{in} \quad 0 \leq \eta \leq f(\xi), \quad 0 \leq \xi < \infty, \label{eq:ss_inner}\\
\begin{rcases} 
      U^{\textrm{out}}_{\xi} + V^{\textrm{out}}_{\eta} = 0, \\
      \lambda(U^{\textrm{out}}_{\xi \xi} + U^{\textrm{out}}_{\eta \eta}) = P^{\textrm{out}}_{\xi} ,\\
      \lambda(V^{\textrm{out}}_{\xi \xi} + V^{\textrm{out}}_{\eta \eta}) = P^{\textrm{out}}_{\eta},
\end{rcases} \quad \text{in} \quad f(\xi) \leq \eta < \infty, \quad 0 \leq \xi < \infty, \label{eq:ss_outer}
\end{gather}
\label{eq:ssNS}
\end{subequations}
to be solved with the no-slip boundary condition $U^{\textrm{in}} = V^{\textrm{in}} = 0$ at the wall, $\eta = 0$, the symmetry condition $U^{\textrm{in}}= V^{\textrm{in}}_{\xi} = U^{\textrm{out}}= V^{\textrm{out}}_{\xi} = 0$ at $\xi = 0$, and the following normal and tangential surface-stress balance, kinematic condition, and continuity of velocities at the as-yet-unknown free surface, $\eta = f(\xi)$, respectively,
\begin{subequations}
\begin{gather}
  (1+f_{\xi}^2)(P^{\textrm{in}}-P^{\textrm{out}})-2(V^{\textrm{in}}_{\eta}-\lambda V^{\textrm{out}}_{\eta})+ \nonumber \label{eq:ssnormal} \\
  2[V^{\textrm{in}}_{\xi} - f_{\xi} U^{\textrm{in}}_{\xi} + U^{\textrm{in}}_{\eta}-\lambda(V^{\textrm{out}}_{\xi} - f_{\xi} U^{\textrm{out}}_{\xi} + U^{\textrm{out}}_{\eta}) ] f_{\xi}=0, \\
  (1- f_{\xi}^2)(V^{\textrm{in}}_{\xi}+U^{\textrm{in}}_{\eta})+2(V^{\textrm{in}}_{\eta}-U^{\textrm{in}}_{\xi})f_{\xi} =\lambda[(1- f_{\xi}^2)(V^{\textrm{out}}_{\xi}+U^{\textrm{out}}_{\eta})+2(V^{\textrm{out}}_{\eta}-U^{\textrm{out}}_{\xi})f_{\xi}],\\
  f/3-\left( \xi/3 + U^{\textrm{in}} \right) f_{\xi} + V^{\textrm{in}} = 0, \label{eq:sskinem}\\
  U^{\textrm{in}} - U^{\textrm{out}} = V^{\textrm{in}} - V^{\textrm{out}} = 0. \label{eq:ssvelocities}
\end{gather} 
\label{eq:ssbcs}
\end{subequations} 
The solution to the elliptic problem~\eqref{eq:ssNS}--\eqref{eq:ssbcs} would determine the asymptotic film shape $f = f(\xi)$ as $\tau \to 0$. Figure~\ref{fig:fig4} shows the results of the numerical integration of~\eqref{eq:ssNS}--\eqref{eq:ssbcs} compared with the time marched film evolution represented in self similar scales, showing a very good agreement. 

For $\lambda\ll 1$, the solution deduced by~\citet{morenobozaetal2020} is seen to be effectively recovered, with an opening angle $\theta_{\textrm{o}} \simeq 37^\circ$~\com{(see Fig.~\ref{fig:fig4})}. Indeed, as seen in~\cite{morenobozaetal2020}, the kinematic condition~\eqref{eq:sskinem} shows that a wedge-shaped far field of the form $f(\xi) = (\xi - \xi_{\textrm{o}}) \tan\theta_{\textrm{o}} $ is compatible with vanishing velocities, valid for $r^2 = \xi^2 + \eta^2 \gg 1$, with $(r, \theta)$ polar coordinates such that $\xi - \xi_{\textrm{o}} = r \cos{\theta}$ and $\eta = r \sin{\theta}$. The opening angle $\theta_{\textrm{o}}$ and the constant $\xi_{\textrm{o}}$ are to be determined numerically. However, insight can be gained by examining the far field on assuming radially decaying velocities with associated stream functions $\psi^{i} \simeq r^{-\ell}$, with $\ell > 0$, such that $V_r^i \sim V_\theta^i \sim r^{-(1+\ell)}$ along rays $\theta = \text{constant}$, where $(V_r^i,V_\theta^i)$ are the associated radial and polar components of the velocity field. On assuming $\psi^{i} = F_{i}(\theta) r^{-\ell}$, the following fourth-order homogeneous problem for $F$ arises:
\begin{figure}
    \centering
    \includegraphics[width=440pt]{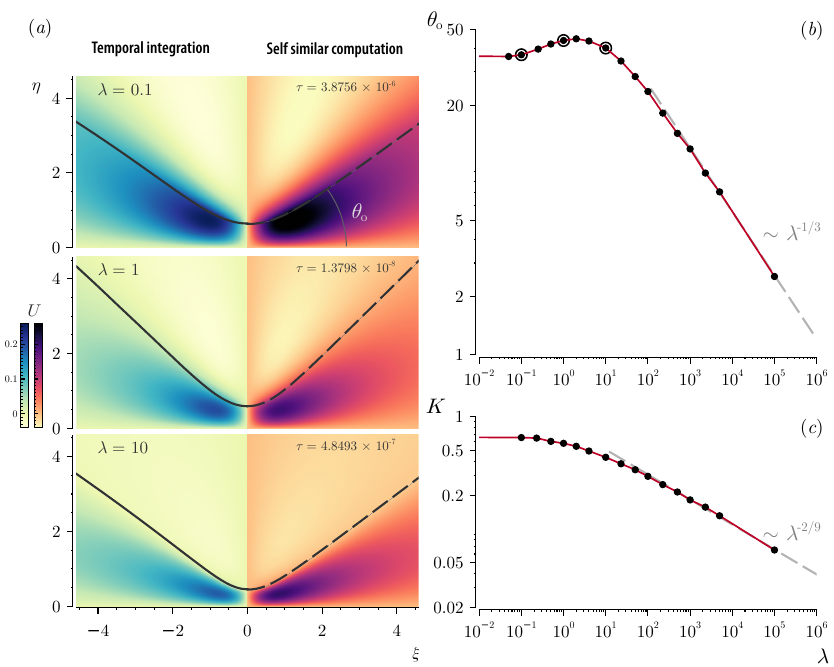}
    \caption{($a$) Self-similar flow structure for $\lambda = 0.1$ (upper panels), $1$ (middle panels) and $100$ (lower panels), for $h_{\textrm{o}} = 10$, illustrated by means of temporal integrations of~\eqref{eq:ns} for $\tau\to 0$ (left panels) and the solution of the self-similar elliptic problem~\eqref{eq:ss_inner}--\eqref{eq:ssvelocities} (right panels). ($b$) Angle and ($c$) prefactor $K$ of the 1/3 power-law as functions of the viscosity ratio, $\lambda$.}
    \label{fig:fig4}
\end{figure}
\begin{equation}
    F_{i}^{\text{(iv)}} + \left[4 + 2\ell (2+\ell) \right] F_{i}'' + \ell^2 (2 + \ell)^2  F_{i} = 0,\label{eq:eqF}
\end{equation} 
to be integrated with the no slip conditions $F_{\textrm{in}}(0) = F'_{\textrm{in}}(0) = 0$, the symmetry conditions $F_i(\pi/2) = F_i''(\pi/2) = 0$, and the normal stress balance $\lambda F_{\textrm{out}}''' - F_{\textrm{in}}''' + \left[ 4 + 3 \ell \left(2 + \ell\right)  \right] \lambda F_{\textrm{out}}' - \left[ 4 + 3 \ell \left(2 + \ell\right)  \right] F'_{\textrm{in}} = 0 $ at $\theta = \theta_{\textrm{o}}$. Since the problem is homogeneous, the system furnished by the prescription of the above boundary conditions for the corresponding constants of integration is also homogeneous. Therefore, the determinant of the so-formed system being zero can be recast into the compatibility condition $\varphi(\theta_{\textrm{o}}, \ell; \lambda) = 0$, where the algebraic function $\varphi$ is omitted here for conciseness. The function $\varphi$ can be solved numerically to give the far-field angle $\theta_{\textrm{o}}$ as a function of the radial decay of the stream functions $\ell$ for given values of $\lambda$, similarly to what was done in~\cite{morenobozaetal2020} for $\lambda = 0$. After retrieving the different values of $\ell$ from the computations in figure~\ref{fig:fig4} one obtains, for instance, $\theta_{\textrm{o}} \simeq 37^\circ$ for $\lambda = 0$ and $\ell \simeq 0.19$, $\theta_{\textrm{o}} \simeq 44.0^\circ$ for $\lambda = 1$ and $\ell \simeq 0.5$, and $\theta_{\textrm{o}} \simeq 19.1^\circ$ for $\lambda = 100$ and $\ell \simeq 0.1$. 

The far-field angle of the interface, $\theta_{\textrm{o}}$, and the prefactor of the $1/3$ power-law, $K$, are plotted as functions of $\lambda$ in Figs.~\ref{fig:fig4}(b) and~\ref{fig:fig4}(c), respectively. Interestingly, the function $\theta_{\textrm{o}}(\lambda)$ presents a global maximum of $44.4^{\circ}$ at an intermediate value of $\lambda=1.9$. It is also deduced that $\theta_{\textrm{o}}\to 37^{\circ}$ as $\lambda\to 0$, consistent with the result of~\citet{morenobozaetal2020}, and that $\theta_{\textrm{o}}$ decreases rapidly for $\lambda\gtrsim 10$, eventually achieving a power-law $\theta_{\textrm{o}}\sim \lambda^{-1/3}$, for $\lambda\gg 1$, demonstrating that the shapes close to rupture become extremely slender for large values of the viscosity ratio.

\section{Concluding remarks}
\label{sec:conclusions}
We have unraveled the influence of an outer immiscible bath on the two-dimensional dewetting dynamics of an ultrathin non-wetting liquid film resting on a solid substrate. A linear stability analysis of the governing Stokes equations has revealed that the most unstable wavelength increases as the bath-to-film viscosity ratio $\lambda$ increases, pointing to the existence of more slender dewetting structures compared to the case without an outer bath, which is recovered as a particular case of our analysis in the limit $\lambda\to 0$. Indeed, nonlinear numerical simulations have confirmed that the local shape of the interface close to rupture becomes increasingly slender as $\lambda$ is increased, and have revealed that the film thinning process is ultimately governed by the same self-similar dynamics as that discovered by~\citet{morenobozaetal2020} for the case without an outer bath, $\lambda=0$. The self-similar dynamics depends on a balance between the van der Waals and viscous forces, with asymptotically negligible surface tension forces, and provides a minimum film thickness $\hmin$ obeying the power law $\hmin=K(\lambda)\tau^{1/3}$, where $\tau$ is the time remaining until rupture and the prefactor $K$ is a monotonically decreasing function of $\lambda$. We have also revealed that the structure of the near-rupture flow is self-similar, and that the interface is wedge-shaped with an opening angle $\theta_{\textrm{o}}(\lambda)$ which decreases with $\lambda$, in agreement both with the linear stability analysis and with the numerical evidence.

As prospects for future works, we point out the need to perform experiments and three-dimensional numerical simulations of the flow configuration studied herein. In addition, a number of effects of practical relevance, like the slip at the solid substrate~\cite{kargupta2004instability,Munch2005,Fetzer2005,martinez2020effect} or the viscoelasticity of the liquid~\cite{rauscher2005thin,munch2006jeffreys,martinez2021non}, should also be taken into account, since both are known to play an important role in the dewetting of polymeric liquids.


\section*{Acknowledgments}
This research was funded by the Spanish MCIU-Agencia Estatal de Investigaci\'on through project PID2020-115655GB-C22, partly financed through FEDER European funds.

\section*{Data availability}
The data that support the findings of this study are available from the corresponding author upon reasonable request.



\bibliographystyle{plainnat}
\bibliography{biblio}

\end{document}